\documentclass[aps,prl,twocolumn,floatfix]{revtex4}

\usepackage{graphicx}
\usepackage{dcolumn}
\usepackage{bm}

\newcommand{\Tr}{\mbox{Tr}}
\newcommand{\grad}{\mbox{grad}}

\renewcommand{\div}{\mbox{div}}
\newcommand{\Symm}{\mbox{Symm}}
\newcommand{\pdiffl}[2]{\frac{\partial #1}{\partial #2}}
\newcommand{\diffl}[2]{\frac{d #1}{d #2}}

\newcommand{\kB}{k_{\text{B}}}
\newcommand{\rWS}{r_{\text{WS}}}
\newcommand{\vWS}{v_{\text{WS}}}

\newcommand{\thetaE}{\theta_{\text{E}}}
\newcommand{\fTQ}{f_{\text{TQ}}}

\newcommand{\condbf}[1]{{#1}}

\begin{document}


\title{Dislocation-based strength model for high energy density conditions}

\date{June 8, 2021; updates to July 25, 2022 -- LLNL-JRNL-825656} 

\author{Damian C. Swift}
\email{dswift@llnl.gov}
\author{Kazem Alidoost}
\author{Ryan Austin}
\author{Thomas Lockard}
\author{Christine Wu}
\author{Sebastien Hamel}
\author{John E. Klepeis}
\affiliation{%
   Lawrence Livermore National Laboratory,
   7000 East Avenue, Livermore, California 94551, USA
}
\author{Pedro Peralta}
\affiliation{%
   Department of Mechanical and Aerospace Engineering,
   Arizona State University, P.O. Box 876106, Tempe, Arizona 85287, USA
}

\begin{abstract}
We derive a continuum-level plasticity model for polycrystalline materials
in the high energy density regime,
based on a single dislocation density and single mobility mechanism,
with an evolution model for the dislocation density.
The model is formulated explicitly in terms of quantities connected closely 
with equation of state (EOS) theory, in particular the shear modulus and
Einstein temperature, which reduces the number of unconstrained parameters
while increasing the range of applicability.
The least constrained component is the Peierls barrier $E_P$,
which is however accessible by atomistic simulations.
We demonstrate an efficient method to estimate the variation of $E_P$ with
compression, constrained to fit a single flow stress datum.
The formulation for dislocation mobility accounts for some or possibly all
of the stiffening at high strain rates usually attributed to phonon drag.
The configurational energy of the dislocations is accounted for explicitly,
giving a self-consistent calculation of the conversion of plastic work to heat.
The configurational energy is predicted to contribute to the mean pressure,
and may reach several percent in the terapascal range,
which may be significant when inferring scalar EOS data
from dynamic loading experiments.
The bulk elastic shear energy also contributes to the pressure,
but appears to be much smaller.
Although inherently describing the plastic relaxation of elastic strain,
the model can be manipulated to estimate the flow stress as a function
of mass density, temperature, and strain rate, which is convenient to
compare with other models and inferences from experiment.
The deduced flow stress reproduces systematic trends observed in elastic waves
and instability growth experiments,
and makes testable predictions of trends versus material and crystal type
over a wide range of pressure and strain rate.
\end{abstract}


\maketitle

\section{Introduction}
Overwhelmingly most research on plastic flow has been devoted to 
engineering materials and simpler variants such as elemental Fe and Al,
for deformation rates and temperatures occurring in typical applications,
and at pressures around ambient, or between zero and those occurring
in high-pressure static reservoirs, which is essentially the same.
However, strength is important in a much wider range of materials and conditions.
An extreme example is the crust of neutron stars \cite{Horowitz2009},
where fracture causing a break in the crust, leading to magnetic reconnection,
has been proposed as a mechanism for giant gamma ray bursts in magnetars,
which are powerful enough to disturb the Earth's ionosphere
from tens of thousands of light years \cite{Campbell2005}.
At the intersection of science and engineering
are the impact of micrometeoroids on spacecraft, and experiments performed
to investigate material properties at pressures into the terapascal scale
during dynamic loading on nanosecond time scales
\cite{hedexpts}, which are directly relevant to the development of
techniques for inertially-confined thermonuclear fusion \cite{icf}.
As well as the sample materials studied, these experiments usually
involve other components whose strength often affects the conditions
imparted to or inferred in the sample.
Simulations and interpretation of these experiments then must rely
on models of strength that have not been calibrated or validated in
relevant conditions.
Strength models can be deduced theoretically, but the detailed mechanisms
of plastic flow are complicated, and have previously involved
extensive multi-scale computational studies using electronic structure,
molecular dynamics, and dislocation dynamics to calibrate a
continuum-level model suitable for simulations of dynamic loading experiments
\cite{Barton2011,Barton2013}.
\condbf{%
Successful plasticity models have also been developed by considering
physically-plausible contributions for the motion and evolution of dislocations,
and calibrating them primarily to experimental measurements probing
the sensitivity to temperature, pressure, strain rate etc
\cite{Preston2003,Rodriguez2011,Gao2012,Lim2015,Austin2018}.
Dynamic loading experiments are difficult and expensive,
and such models are usually calibrated over a narrow range of
relatively low pressures.
Otherwise, strength models used for such simulations are essentially
empirical descriptions of the flow stress as a function of state,
calibrated to much lower pressures, temperatures, and strain rates,
with minor empirical or guessed modifications.
}
One notable {\it ad hoc} simplification is that, in all widely-used
continuum models, the heat imparted by flow against strength is assumed to be a
constant fraction of the plastic work, often simply unity. 
In contrast,
experimental measurements \cite{Mason1994,Kapoor1998,Soares2021}
and theoretical studies \cite{Rosakis2000,Hodowany2000,Zubelewicz2008,Zubelewicz2019}
have found that the fraction may vary between 0.1 or less to $\sim$1
over the course of a single loading event.

We have recently developed methods to predict several properties of matter
efficiently over a wide range of states,
including the equation of state (EOS) \cite{Swift_ajeos_2019,Swift_wdeos_2020},
the asymptotic freedom of ions leading to the drop in their
heat capacity from 3 to $3/2\,\kB$ at high temperature \cite{Swift_iontherm_2020},
and also the shear modulus \cite{Swift_ajshear_2021},
which is the driving force behind plastic flow.
We previously developed a dislocation-based crystal plasticity model
for deformation at high pressures and strain rates,
formulated with the intention of requiring less computational effort to
calibrate \cite{Swift_xtplast_2008}.
In the work reported here, we extend the concepts behind this model,
informed by recent developments and experience,
and combine it with our new predictions of shear modulus to produce
a strength model valid potentially over the full range of normal matter,
with particular interest in conditions occurring in high energy density (HED)
applications such as loading induced by laser ablation.

We do not expect the model described here to be as accurate 
as models constructed from more detailed multi-scale studies over a more
restricted range of states \cite{Barton2011,Barton2013},
\condbf{or by considerating detailed energetics and kinematics
of a dislocation, such as the bowing geometry, pinning, and cross-slip
\cite{Lim2015,Hunter2015,Blaschke2020,Hunter2022}}.
However, the present model is significantly simpler to implement \condbf{and calibrate},
includes other effects neglected in previous strength models,
and can be applied in arbitrarily extreme conditions
without further extension or calibration.
Although we use a particular electronic structure method
to supply properties needed to infer the strength over a wide range of states,
the plasticity model is constructed to make it straightforward to substitute
properties calculated using more detailed or accurate methods where available.
It is preferable to use a strength model constructed using
potentially inaccurate components than one constructed to be accurate
under irrelevant conditions that would definitely be inaccurate
outwith its range of calibration, or no model at all.
This approach may also be suitable to estimate the plastic flow rate
associated with twinning, as this process also fundamentally involves
atoms hopping past a Peierls barrier.

\section{Dislocation dynamics}
In contrast to our previous study \cite{Swift_xtplast_2008}, we focus here
on plastic flow in polycrystalline aggregates, omitting some effects
such as the change in resolved shear stress from finite elastic strains,
and employing implicit averages over crystal orientations and slip systems.
The constitutive equations can be solved in different ways,
influenced by the internal structure of simulation programs in which they
are implemented.
As before \cite{Swift_xtplast_2008},
the intended method of solution is in continuum dynamics simulations 
in which the local material state includes the elastic strain tensor,
from which the stress tensor $\tau$ is calculated using the
bulk and shear moduli $B$ and $G$,
properties of the material state and in particular of the mass density
$\rho$. $G$ can be predicted from electronic structure theory,
as in our recent wide-range study \cite{Swift_ajshear_2021}.
The driving force for plastic deformation is the deviatoric stress
\begin{equation}
\sigma\equiv\tau+Ip,
\end{equation}
where the mean pressure $p=-\Tr\,\tau/3$.
The deviatoric stress $\sigma=G\epsilon_e$ 
where $\epsilon_e$ is the elastic strain deviator
and $G$ the shear modulus.
In finite-strain continuum mechanics,
the isotropic and shear strain rates are the isotropic and deviatoric
decomposition of the symmetric part of the gradient tensor
of the material velocity field $\vec u(\vec r)$:
\begin{equation}
\dot\epsilon\equiv\Symm(\grad\,\vec u)+\dot\mu I,
\end{equation}
the antisymmetric part describing rotation.
The isotropic strain rate gives the rate of compression,
$\dot\mu\equiv-\div\,\vec u=\dot\rho/\rho$.
The deviatoric strain rate $\dot\epsilon$ is the sum of elastic and plastic
components.
Here we use dislocation dynamics to determine the plastic strain rate.
For polycrystal-averaged materials with shear stress calculated 
using a single, isotropic shear modulus, 
the scalar plastic strain rate $\dot\epsilon_p$
acts to reduce the elastic strain by radial relaxation toward the state
of isotropic pressure.
This approach is standard for continuum simulations of the response of
matter to dynamic loading \cite{Benson1992}.

The plastic strain rate $\dot\epsilon_p$ is related to the dislocation density
$\rho_d$ and the mean speed of dislocation motion $\bar v_d$
by a modified Orowan equation \cite{Barton2013}
\begin{equation}
\dot\epsilon_p=\frac\eta M \rho_d b \bar v_d
\end{equation}
where $b$ is the magnitude of the Burgers vector, $M$ the Taylor factor
representing an effective number of active slip systems, and
$\eta$ a microstructure factor representing the number of dislocation types
(1 for screw only, 2 for screw and edge).
$\bar v_d$ can be obtained from a kinetics model, giving a hopping rate $Z$, so
$\bar v_d=b Z$.
$\rho_d$ is usually defined in conventional metallurgical terms as the
dislocation line length per volume of material.
At finite compressions, $\rho_d$ and $b$ scale with the mass density $\rho$
as $\rho^{2/3}$ and $\rho^{-1/3}$ respectively.
The constitutive equations become simpler, and more straightforward to
generalize to large changes in compression, if expressed instead
using the fraction of atoms on a dislocation,
\begin{equation}
\phi_d=\rho_d \frac{m_a}{b\rho}
\end{equation}
where $m_a$ is the mass of an atom.
Then
\begin{equation}
\dot\epsilon_p = \frac \eta M \phi_d Z \gamma
\quad:\quad \gamma\equiv\frac{6f_v^3}\pi
\end{equation}
The ratio of Burgers vector to Wigner-Seitz radius, $b/\rWS$,
is constant for a given crystal structure
(Table~\ref{tab:fv});
for convenience we consider the distance from the equilibrium position
to the peak of the Peierls barrier, $b/2$, as $f_v\equiv b/2 \rWS$.

\begin{table}
\caption{Scaling between Burgers vector and Wigner-Seitz radius,
   for simple crystal structures.}
\label{tab:fv}
\begin{center}
\begin{tabular}{|l|l|}\hline
{\bf structure} & $f_v$ \\ \hline
body-centered cubic (bcc) & $\frac 12\left(9\pi\right)^{1/3}$ \\
face-centered cubic (fcc) & $\frac 1{\sqrt 2}\left(2\pi/3\right)^{1/3}$ \\
hexagonal close-backed (hcp), basal & $\left(2\pi/3\right)^{1/3}$ \\
diamond & $\frac 1{\sqrt 2}\left(4\pi/3\right)^{1/3}$ \\
\hline\end{tabular}
\end{center}
\end{table}

Previous dislocation-based strength models were constructed,
and $\bar v_d$ calculated, assuming dislocation motion was impeded
by barriers to `forward' motion
\cite{Preston2003,Barton2011,Barton2013,Austin2018}.
The resulting Arrhenius hopping probability gives a non-zero $\bar v_d$
at zero applied stress, and various modifications have been employed
to ensure that $\bar v\rightarrow 0$ as $\|\sigma\|\rightarrow 0$,
including the use of an error function instead of an Arrhenius rate
\cite{Preston2003}, and a minimum stress below which $\bar v_d$ was set to zero
\cite{Barton2011,Barton2013,Austin2018}.
As we pointed out previously \cite{Swift_xtplast_2008}, this problem is avoided by recognizing that
atoms have no inherently preferred hopping direction, but the local stress
field biases hops in the direction that reduces shear stress,
because the stress reduces the Peierls barrier in that
direction and increases it in the opposite direction. 
The resulting hopping rate is
\begin{equation}
Z=Z_0\left[\exp\left(-N\frac{E_P-E_\tau}{\kB T}\right)
          -\exp\left(-N\frac{E_P+E_\tau}{\kB T}\right)\right]
\label{eq:rate}
\end{equation}
where $Z_0$ is the attempt rate, $E_P$ the Peierls barrier,
$N$ the number of coordinated atom jumps required for the dislocation to move,
$T$ the temperature, and $\kB$ Boltzmann's constant.
$E_\tau$ is the effect of the local stress on the forward and reverse
Peierls barriers, $\|\sigma\| \vWS f_v$
\condbf{where $\vWS$ is the Wigner-Seitz volume}.
As well as being numerically-better conditioned as $\sigma\rightarrow 0$,
the inclusion of reverse hopping reduces the net rate at elevated temperatures,
giving similar phenomenological behavior as has been attributed to
phonon damping, but without an additional explicit contribution or parameters.
The barrier-lowering effect of the applied stress for forward hopping has
also been considered by other researchers \cite{Deo2005},
and the effect of reverse hopping has been included in dislocation based
studies of creep \cite{Nichols1971}.

The Arrhenius rate itself represents the probability of an atom
following a Maxwell-Boltzmann velocity distribution crossing a positive
energy barrier greater than $\kB T$.
If the stress is high enough for the barrier to become substantially smaller, 
the Arrhenius rate is incorrect; it should certainly not exceed unity.
The mean dislocation speed is thus limited to $Z_0 b$ at high stresses,
\condbf{just as the `relativistic' effect attributed to phonon drag}.

At the microstructural level, stress is resolved on the specific slip systems  
and stress relaxation occurs by the motion of dislocations of each Burgers vector,
with separate kinetics for each.
The resolved shear stress and Burgers vector occur in the exponent of the
rate equation Eq.~\ref{eq:rate},
so in principle the response of polycrystalline material may be non-linear
with respect to averaging of the resolved stress in the exponent and
averaging of the plastic strain rate:
a potential source of inaccuracy in predicting the flow stress.
This averaging could be performed explicitly for a given crystal structure
to go beyond the use of an assumed Taylor factor,
and multiple contributions to the rate could be included with
different Peierls barriers and effective attempt rates, to account
for non-Schmid effects even in an average polycrystalline model.
For now, we simply take a constant value $M$ appropriate for the 
crystal structure and material.

Similarly, there are in general several barriers that affect the dislocation mobility,
including the Peierls barrier for the ideal lattice,
impurities such as interstitials, and intersections with dislocations
of different Burgers vector.
Usually one or other barrier dominates and controls the plastic strain rate,
depending on the material, compression, temperature, applied stress,
and microstructural state including the instantaneous dislocation density.
In principle, several types of barrier could be incorporated even in
a polycrystal-averaged model,
potentially including competition between multiple slip systems
leading to non-Schmid effects and a varying effective Taylor factor.
In the present work, we consider a single
barrier and investigate its applicability over a wide range of conditions
in the HED regime.



\section{Evolution of dislocation density}
The model so far gives the plastic strain rate for a given dislocation density.
Predictions of the rate of change of shear stress can be made given
a measurement of the initial dislocation density in a material,
but the dislocation density evolves as plastic flow occurs.
A key theoretical insight leading to dislocation-based plasticity models
has been the prediction that, at a given strain rate,
the dislocation density evolves rapidly toward an equilibrium value
\cite{Barton2011}, which can be estimated from large-scale dislocation dynamics
simulations and parameterized as part of the strength model.
Here we investigate an approach in which the equilibrium density $\hat\phi_d$ is
an emergent property rather than a separately-calibrated definition.
Ignoring the spontaneous generation of dislocations,
if dislocations are sufficiently far apart and thus unable to annihilate
each other, 
and considering dislocation loops from Frank-Read type sources,
the rate of dislocation generation is proportional to the rate of dislocation motion:
\begin{equation}
\dot\phi_d=\frac\pi 2\frac{\phi_d}M Z=\frac\pi 2\frac{\dot\epsilon_p}\gamma.
\end{equation}
For a given slip system, if dislocations of opposite Burgers vector
occur in equal numbers, one contribution to their annihilation is
collision as they move in opposite directions under the applied stress.
In addition, dislocations of opposite sign attract each other through
their mutual elastic strain fields, which decay with distance from
each dislocation as $Gb/r$.
Using the same method for predicting the hopping rate as for the applied stress,
\begin{equation}
Z_a=Z_0\left[\exp\left(-N\frac{E_P-E_m}{\kB T}\right)
            -\exp\left(-N\frac{E_P+E_m}{\kB T}\right)\right]
\end{equation}
where $E_m$ is the adjustment to the Peierls barrier from the mutual interaction
field of the dislocations,
\begin{equation}
E_m=\frac{G \vWS f_v}{\bar L}
\end{equation}
where $\bar L$ is the mean separation between dislocations of opposite sign,
in units of the Burgers vector,
\begin{equation}
\bar L=\sqrt{\frac{2 M}{\phi_d}}.
\end{equation}
The net rate of change in dislocation density is then
\begin{equation}
\dot\phi_d=\frac\pi 2\frac{\phi_d}MZ\left(1-\frac 2{\bar L}\right)
   -\frac{2\phi_d Z_a}{\bar L}
=\frac{\dot\epsilon_p}\gamma \frac\pi 2
\left(1-\frac 2{\bar L}\right)
   -\frac{2\phi_d Z_a}{\bar L}.
\label{eq:dotphid}
\end{equation}
Adjustment factors may be included for each term to match more detailed
modeling where available.

If the dislocation population is in equilibrium, $\dot\phi_d=0$,
and $\hat\phi_d$ can be deduced from Eq.~\ref{eq:dotphid},
which can be solved numerically by bisection since $0\le\phi_d\le 1$.
Dependence on state and strain rate enters through the attraction term only.
Neglecting this term,
\begin{equation}
\hat\phi_d=M/2 > 1.
\end{equation}
This limiting case could be the result of over-simplification in the
generation and collision terms, such as the use of the mean separation $\bar L$.
There presumably is a limiting value of $\phi_d$, $\phi_{\text{max}}$ say,
above which the concept of dislocations is meaningless.
In practice, $\phi_d$ must remain well below 1
for the crystal lattice to be stable \cite{Cotterill1977}.
To correct for too large a $\phi_{\text{max}}$ from the asymptotic limit
of the generation and collision terms, we include an additional
saturation term to limit $\phi_d$ to some chosen $\phi_{\text{max}}$:
\begin{equation}
\dot\phi_d=R_m(\phi_{\text{max}}-\phi_d)-R_c-R_a
\end{equation}
where $R_m$, $R_c$, and $R_a$ are the rates from multiplication, collision,
and attraction respectively, from Eq.~\ref{eq:dotphid}.

As we are neglecting homogeneous thermal nucleation of dislocations,
the equilibrium dislocation density at static conditions 
$\hat\phi_d(\dot\epsilon_p=0)$ is zero.
This is physically reasonable as the dislocation population in virtually all metals
as prepared is metastable under ambient conditions.
In practice, we can specify a reasonable value for the initial dislocation
density, and find that evolution toward zero is negligible
under ambient conditions on the time scales of HED experiments.
The estimates of $\phi_d$ {\it in-situ} during dynamic deformation
could be tested by experiments e.g. using synchrotron x-ray measurements.

Dislocation densities in bulk engineering materials
are typically $\sim 10^8$/cm$^2$ i.e. $\phi_d\sim 10^{-7}$.
For any reasonable value of $f_b$, the resulting flow stress was much
larger than observed in dynamic loading experiments, indicating that $\phi_d$
must increase rapidly to accommodate the resulting plastic strain rate.
Such increases have been inferred experimentally in studies of shaped charge
jets \cite{Goldstein1993}, in molecular dynamics simulations
\cite{Srinivasan2006}, and in dislocation dynamics simulations
\cite{Barton2011,Barton2013} of high-rate loading.
Except for transient effects as the strain rate is varied,
it has been found possible to estimate the
flow stress during high-rate loading from the equilibrium dislocation
density instead of considering its instantaneous evolution \cite{Rudd2019}.
In the figures below, we do the same and calculate the equilibrium density
$\hat\phi_d$ given the state and $\dot\epsilon_p$.

A corollary is that it may be simpler than often suspected to predict the
strength of a material following a phase change.
Dislocations in the original phase are likely to disappear in any diffusive
phase transformation, and the
dislocation population in the new phase is likely to depend on the details of 
crystal nucleation and growth.
However, if the dislocation density evolves rapidly toward an equilibrium 
value as further deformation occurs, the density on first formation may matter
little for the subsequent flow stress.

\section{Strain and grain hardening}
Dislocation-based strength models have previously included
a semi-empirical term to represent the increase
of flow stress with plastic strain,
interpreted as associated with a resulting increase in $\rho_d$,
for instance \cite{Barton2013} $\hat\tau=\zeta b G\sqrt{\rho_d}$.
For our purposes it would be preferable to represent this kind of
behavior consistently with the mobility of the dislocations or evolution of 
$\phi_d$.
Strain hardening could be interpreted in several related ways.
When dislocations intersect and tangle,
a greater bowed length of dislocation is needed to accommodate a given
amount of plastic strain.
In a bowed dislocation, a greater proportion of the length must overcome
a barrier closer to the full $E_P$ rather than $E_P-E_\tau$,
which could be represented a hop rate $Z$ depending on the mean free
dislocation length $\bar L$.
Alternatively, this effect could be represented as a reduction in $\epsilon_p$
for a given $\dot\phi_d$, and thus an increase in the strain energy of the
dislocations $e_d$, i.e. $\dot\epsilon(\bar L)$ or $M(\bar L)$.
An additional barrier term could be included for unpinning of tangled dislocations.
However it is represented, strain hardening may also have an effect on
dislocation annihilation.

As a trial model of hardening, we estimate the
mean distance between dissimilar dislocation intersections as
\begin{equation}
\bar L' b\quad:\quad\bar L'=\sqrt{\frac{M-1}{\phi_d}},
\label{eq:barnp}
\end{equation}
leading to a (further) modified Orowan equation
\begin{equation}
\dot\epsilon_p=\frac\eta M \phi_d Z \gamma f_h
   \quad:\quad f_h\equiv 1-\frac 1{\bar L'}.
\end{equation}
The hardening term $f_h$ can thus be considered equivalent
to an effective $M$, $Z$, or $\phi_d$.
A further benefit is that we can attempt to incorporate 
the Hall-Petch grain size effect
\cite{Hall1951,Petch1953} in the same spirit
by including an additional contribution to $\bar L'$
representing the stiffening effect of grain boundaries:
\begin{equation}
f_h=1-\sqrt{\frac{\phi_d}{M-1}+f_{\text{HP}}\phi_g},
\end{equation}
where $\phi_g$ is, in similar convention to $\phi_d$, the fraction of atoms
on a grain boundary.
Considering the surface area to volume ratio of a grain, expressed in terms
of $\rWS$,
\begin{equation}
\phi_g\simeq\frac{12\rWS}{\lambda_g}
\end{equation}
for equiaxed grains.
$f_{\text{HP}}$ is a geometrical factor for the distance from a point on
a dislocation to the nearest grain boundary.
We estimate $f_{\text{HP}}$ to be approximately 2,
though if based on the average chord of a sphere \cite{Dirac1943}
it may be closer to 3.
In practice we limit $f_h\ge 0$ for numerical safety.

\section{Energetics}
An important unresolved question in plasticity is the relation between 
plastic work and heating.
Virtually all strength models in dynamic loading simulations assume that
a constant fraction of plastic work appears as heat,
be represented by the Taylor-Quinney factor,
$\fTQ$ \cite{Farren1925,Taylor1934}.
In many implementations, {\it all} plastic work is converted to heat,
i.e. $\fTQ=1$.
Experimentally, $\fTQ$ has been observed to vary during the loading history
\cite{Mason1994,Rosakis2000,Hodowany2000,Soares2021}.
It may be close to zero early on, subsequently rising close to or even
exceeding unity, though other behaviors have been observed and the
measurement is difficult.
We hope to capture such behavior by accounting for the
configurational energy of the dislocation population.

The core energy of a screw dislocation amounts to
around $Gb^2/2$ per length, i.e. $Gb^3/2$ per $b$ of line.
The elastic strain energy for well-separated dislocations is similar.
The specific energy associated with the dislocation population is then
\begin{equation}
e_d\simeq G\frac{f_v^3\phi_d}\rho.
\label{eq:e_d}
\end{equation}
At high dislocation densities, the strain fields from dislocations
of opposite Burgers vector tend to cancel out, leading to 
a logarithmic correction to the energy from the finite integration
limit in radius.
Molecular dynamics (MD) simulations of the energy of a population of 
dislocations have been found to follow a dependence like
\begin{equation}
e_d\simeq\frac{Gb^2}{4\pi}\frac{\rho_d}\rho\ln\frac 1{b\sqrt{\rho_d}},
\end{equation}
depending in detail on the distribution of dislocation types
and the grain orientation with respect to the load, and subject to statistical
scatter \cite{Stimac2022}. 
To capture this behavior in a bounded way, preserving the core energy
at high dislocation densities, we make the replacement
\begin{equation}
\phi_d\rightarrow\alpha\phi_d\ln\sqrt{\frac{\phi_c}{\phi_d}}
\end{equation}
in Eq.~\ref{eq:e_d},
where
\begin{equation}
\frac 1\alpha=\ln\frac{\phi_{\text{max}}}{\phi_{\text{ref}}}\simeq 4\pi,
\phi_c=\phi_{\text{max}}e^{1/\alpha}.
\end{equation}
Potential energy stored in the dislocation population in this way
reduces the energy available for plastic heating.
Conversely, as dislocations annihilate, which may occur at zero total strain
rate, the temperature will rise.
Considering the total working rate against shear strain $\dot e_s$
as comprising contributions from the elastic shear energy $\dot e_e$ and
the change in dislocation density $\dot e_d$ \cite{Heighway2019}, 
we can calculate 
a self-consistent plastic heating rate $\dot e_h=\dot e_s-\dot e_e-\dot e_p$
and thus a self-consistent, instantaneous TQ factor
$\fTQ=\dot e_h/(\dot e_s-\dot e_e)$.

If the material is compressed isotropically
at a constant dislocation population $\phi_d$,
the specific energy of the dislocations changes.
Neglecting entropy effects, this change appears as a contribution to the
pressure
\begin{equation}
p_d=\rho e_d\left(\frac\rho G \pdiffl G\rho-1\right)
\label{eq:pd}
\end{equation}
(see Appendix A).
In principle, $p_d$ could have either sign,
and it is notable that a constant $G$ would give a negative contribution to pressure.
In our atom-in-jellium predictions of $G(\rho)$ \cite{Swift_ajshear_2021},
the effective exponent of $\rho$ (discussed in Appendix A) 
was almost always greater than 1, so $p_d>0$.
In MD simulations of plastic flow at constant pressure
\cite{Zepeda-Ruiz2017},
an expansion of 0.5 to 1\,$\rWS$ per $b$ of dislocation length has been
observed \cite{Bulatov2021}.
Counteracting this expansion to maintain a constant mass density 
corresponds to a similar increase in pressure.


Our analysis ignores any dilatation caused directly by the dislocations,
by less efficient packing of atoms, which would be in addition to any
contribution from the density-dependence of the shear modulus.
For non-close-packed structures, it is possible that such dilatation may not
be significant, but one might expect it to contribute in close-packed
structures.
The MD simulations we compared with should include both effects, but
the scatter in the reported density change made it impossible to
separate these contributions.
It would be straightforward to include such a geometrical dilatation term,
but this would require an extra material parameter depending at least on mass density,
and so we did not consider it further in the present work.

If mean pressure depends on $\phi_d$,
it follows that the initial state in a simulation should be consistent with 
the initial dislocation density: the initial mass density should be adjusted
to give the correct initial pressure at that temperature and dislocation density.
As discussed above, the value of $p_e$ we found for relevant initial values
of $\phi_d$ was small enough to be negligible for applications investigated
so far, though this adjustment should be made in principle.


\section{Finite elastic strain}
In continuum dynamics programs intended for high pressure simulations,
the mean pressure is taken to depend on isotropic strain only,
through the EOS $p(\rho,T)$, and the resulting stress deviator
$\sigma\equiv\tau+pI$ is assumed to be traceless.
This is only correct if the EOS accounts for the whole of the isotropic stress.
We found previously using electronic structure simulations of
finite elastic shear
deformations \cite{Swift2003} that this assumption is incorrect,
i.e. that the mean pressure at a given mass density and temperature
varies with the elastic strain deviator.
This effect has been observed in fused silica,
and predicted to occur in metals and other ceramics \cite{Scheidler1996}.
We assessed the effect as probably unimportant on microsecond and longer
time scales, where elastic strains in metals do not typically exceed
$\sim 10^{-3}$, but noted that it may matter on shorter time scales such as
are explored in laser-loading experiments, where elastic strains may reach
several percent.

Thermodynamically, any contribution to mean pressure from the strain deviator
should be caused by the elastic distortional energy,
\begin{equation}
e_e=\frac 1{2\rho}G\|\epsilon_e\|^2.
\end{equation}
Neglecting the temperature dependence of $G$, and at constant $\|\epsilon_e\|$,
the contribution to mean pressure is (again per Appendix A)
\begin{equation}
p_e=\rho e_e\left(\frac\rho G \pdiffl G\rho-1\right).
\label{eq:pe}
\end{equation}
See Appendix B for further discussion.


\section{Flow stress}
Although inherently describing the dependence of the plastic strain rate 
on the applied stress,
a flow stress can be deduced from this model.
The flow stress is useful for comparison with experiment, or in
simulations where it is not practical to implement the plastic strain rate
as a relaxation of elastic strain.
Consider a situation where the total strain rate is $\dot\epsilon$
at some $\rho$ and $T$.
The flow stress $Y$ is the external stress needed to give 
$\dot\epsilon_p\simeq\dot\epsilon$.
If the dislocation population is in equilibrium, $\dot\phi_d=0$,
and $\phi_d$ can be deduced from Eq.~\ref{eq:dotphid}.
The shear stress $\|\sigma\|=Y$ can then be found such that
\begin{equation}
\frac\eta M \phi_d Z(\sigma)\gamma f_h=\dot\epsilon_p
\end{equation}
i.e.
\begin{equation}
Y=\sqrt{\frac 32}\frac{\kB T}{\vWS f_v}\sinh^{-1}\left[
\frac{\dot\epsilon_p}{2\gamma\phi_d f_h}\frac M\eta\frac{\exp(E_P/\kB T)}{Z_0}\right].
\label{eq:yeff}
\end{equation}
At the equilibrium dislocation density, the leading term in dislocation density
is roughly proportional to $\sqrt{\phi_d}$, consistent with Taylor hardening
\cite{Taylor1934a} and observations from molecular dynamics simulations
of plasticity \cite{Stimac2022}.

Eq.~\ref{eq:yeff} can be used to construct an approximate model
expressing for example $Y(\rho,\dot\epsilon_p)$ along some reference curve
such as an isentrope.
To estimate the flow stress away from the reference curve, i.e. at different
temperatures, Eq.~\ref{eq:yeff} can be differentiated with respect to
temperature to obtain a softening relation
\begin{equation}
\left.\pdiffl YT\right|_\rho=\frac YT-\frac{E_P}{\vWS f_v T}
\frac \xi{\sqrt{1+\xi^2}}
\end{equation}
where
\begin{equation}
\xi\equiv\frac{\dot\epsilon_p}{2\gamma\phi_d f_h}\frac M\eta\frac{\exp(E_P/\kB T)}{Z_0}
\end{equation}
Along an isochore, $\partial Y/\partial T$ may be positive,
i.e. the material may harden rather than soften with temperature,
depending on the relative magnitude of the two terms.

The flow stress $Y$ above is not a yield stress:
$\dot\epsilon_p>0$ for stresses $\|\sigma\|<Y$.
However, $Y$ provides an indication of when the material responds to
an externally-applied stress largely with plastic flow
rather than elastic strain.

\section{Properties from electronic structure}
$Z_0$ can be determined from the ion-thermal contribution to the EOS.
Used in conjunction with EOS constructed using atom-in-jellium theory,
it is convenient to use the Einstein temperature $\thetaE$,
which is calculated by perturbation of the nucleus from its equilibrium
position.
The corresponding Einstein vibration frequency 
\begin{equation}
Z_E=\kB \thetaE/h,
\end{equation}
where $h$ is Planck's constant.
In the absence of more detailed information,
the Peierls barrier can also be estimated from $\thetaE$,
assuming similarity of the shape of the potential surface experienced by 
the atoms.
$Z_E$ is calculated assuming a harmonic potential, with a stiffness
\begin{equation}
k=m_a\left(2\pi Z_E\right)^2.
\end{equation}
The potential surface is stiffer than harmonic along directions toward
a neighboring atom, and softer along other directions.
Along a Burgers vector, the force must become zero at $b/2$.
We estimate $E_P$ by extrapolating the harmonic potential at some fraction 
of $b$, equivalent to some fraction $f_b$ $O(1)$ of $\rWS$:
\begin{equation}
E_P\simeq k\left(f_b \rWS\right)^2.
\end{equation}
Suggestively, approximately correct flow stresses are obtained with
$f_b\sim 0.1-0.3$.
This range is consistent with our generalization
of the Lindemann melting law which matches melting of the one-component plasma
with $f_b=0.15$, and melting at lower pressures with similar or slightly lower
values \cite{Swift_wdeos_2020,Swift_ajmelt_2020}.
This similarity suggests a somewhat different connection with melting than
previous suggestions of a dislocation-mediated process \cite{Stacey1977}
or critical equilibrium population of dislocations \cite{Burakovsky2000}.
The Lindemann parameter is usually interpreted as a melting criterion
involving the mean square displacement of the atoms.
Our observation suggests that this mean square displacement may correspond
to the kinetic energy required for atoms to hop past the Peierls barrier,
and thus to an abrupt decrease in the resistance to shear stress.

The Einstein or Debye frequencies are average phonon frequencies
that seem reasonable to represent a typical attempt rate for an atom to
hop past a potential barrier, at least to predict systematic trends.
Phonon frequencies range from zero to a value corresponding to the
stiffest elastic modulus.
Low frequencies represent acoustic modes in which atoms move in the same
direction as their neighbors.
High frequencies represent optic modes in which atoms oscillate in antiphase
with their nearest neighbors.
Neither leads to hopping.
In the atom-in-jellium model, surrounding atoms are treated as uniform
positive and negative charge densities, in which neighbors and gaps are
not distinguished.

Our previous melting studies used the Debye temperature inferred from the
jellium oscillation model \cite{Swift_wdeos_2020,Swift_ajmelt_2020}.
Here we use the Einstein temperature, for a more direct connection with
the stiffness of the effective interatomic potential.
Because of the method used to infer Debye from Einstein temperature
in our previous work, $f_b=0.15$ in the melting studies requires an additional
scaling factor $\sqrt[3]{\pi/6}\simeq 0.8$ for use with the Einstein temperature.
Accordingly, we consider a default value of 0.12.

\section{Performance}
The most direct predictions and comparisons of a plasticity model
are the observables in a specific loading experiment.
However, experimental measurements on dynamic loading have wide variations
in loading history,
and require an assessment of the EOS and shear modulus as well as the
plastic flow,
so we will report experiment-by-experiment comparisons elsewhere.
Some trends are observed repeatedly.
The flow stress inferred from elastic precursor waves
is typically several times greater on nanosecond time scales typical of
laser ablation experiments than on microsecond time scales typical of
gun and high explosive experiments \cite{laservsgun}.
In contrast, the flow stress inferred from 
Rayleigh-Taylor strength experiments at pressures in the $\sim 100$\,GPa
range is typically a couple of times higher than predicted using the
Steinberg-Guinan (SG) strength model \cite{rtstren}.

For illustration here, we predict systematic trends in the plastic
response of several elemental metals of different crystal structures,
whose strength is of current interest in high pressure experiments.
The predictions were made using Eq.~\ref{eq:yeff},
but were verified at representative states by simulation of 
a single homogeneous material element subjected to a constant strain rate
(i.e. with the dislocation density allowed to evolve) and also with
hydrocode simulations with a relevant applied loading history,
in which the strain rate evolves in position and time as the loading wave
propagates and evolves through the sample, and the dislocation density
evolves continually along with the local state.

\subsection{Calibration of plasticity models}
To construct plasticity models, we use our previous
atom-in-jellium predictions of the EOS and shear modulus
\cite{Lockard_highZ_2021,Swift_ajshear_2021}.
For simplicity,
we ignore the effect of high-pressure phase transitions,
though these can be taken into account.

\begin{figure}
\includegraphics[scale=0.72]{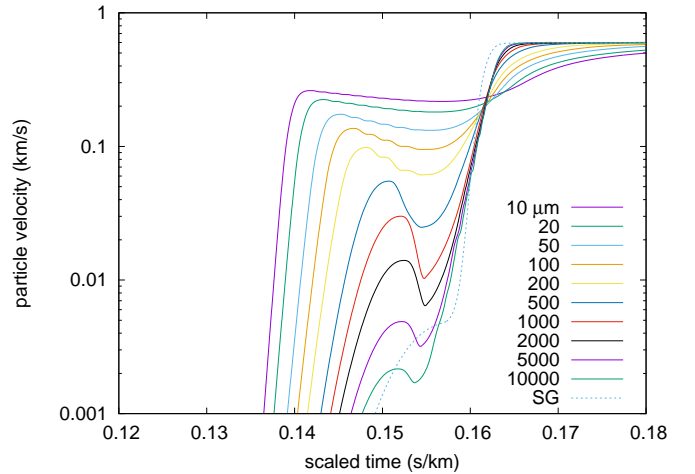}
\caption{\condbf{%
Particle velocity history calculated after shock wave has propagated through
   different thicknesses of Al using the dislocation model,
   compared with simulation using SG model used to
   calibrate $f_b$ assuming a strain rate of $10^6$/s.
   The drive pressure is 10\,GPa in all cases.
   Scaled time (scaled by the propagation distance) is shown to capture the
   response over the wide range considered.
   Each trace shows an elastic precursor wave followed by plastic structure
   before the peak velocity is reached.
   The SG model is rate-independent and therefore scale-invariant.
   With the dislocation model, breakout of the elastic wave is earlier than
   with the SG model because the amplitude is much higher near the driven surface
   and propagates faster.}
}
\label{fig:Alshocklen}
\end{figure}

For each element, we adopted $f_v$ for the appropriate crystal structure,
and took a value of $M$ typical for elements of that structure.
We chose $\eta=1$ on the basis that, for large strains, any initial population
of pure edge dislocations would be swept away, leaving dislocations produced by
Frank-Read sources, as discussed above.
The sole remaining parameter is then $f_b$.
Because of its connection to melting of the one-component plasma, we
first chose $f_b=0.12$ to give an entirely first-principles
(but typically inaccurate) strength model.
We also determined a value $f_b$ to reproduce an experimental measurement
or inference of flow stress.
This could be done by performing simulations of a specific experiment, such
as of an elastic wave amplitude.
For convenience and consistency between different materials,
we instead used the SG value of 
ambient yield stress $Y_0$ for each material
\cite{Guinan1974,Steinberg1980,Steinberg1993}.
This parameter was determined from the elastic wave amplitude observed
in gas gun or explosive-driven measurements on centimeter-scale samples,
and thus represents a curated ensemble of experiments on the response 
to dynamic loading at low pressure.
The SG parameters are quoted without uncertainties, so we report the
corresponding $f_b$ values to high precision (Table~\ref{tab:params}),
in order to reproduce the SG flow stress to at least two significant figures.
We used Eq.~\ref{eq:yeff},
made the common assumption of a representative strain rate,
and took the equilibrium dislocation density $\hat\phi_d$ for that rate.
Values of the representative strain rate for experiments
on millimeter to centimeter scale samples
measuring the response on microsecond scales
are commonly held to be $\sim 10^5 - 10^6$/s.
We determined values for $f_b$ assuming different strain rates,
and found that elastic wave amplitudes predicted using the SG model
were reproduced closely in centimeter scale samples when calibrated 
using $10^6$/s, i.e. the present model and calibration process is consistent 
with a characteristic strain rate of $10^6$/s for elastic precursor waves
in centimeter-scale samples (Fig.~\ref{fig:Alshocklen}).
This is the normalization used for all results shown in the present work.

\condbf{%
The fitted value of $f_b$ for each material considered is shown in
Table~\ref{tab:params}.
The complete parameter set for each, including $\thetaE(\rho)$ deduced
from the atom-in-jellium calculations, is available online at
\cite{Swift_ajdddata_2022}.
}

The figures below also show the SG maximum work-hardening from the 
$Y_{\text{max}}$ parameter,
which represents the maximum yield stress reported for each metal
at ambient pressure and temperature.
The present dislocation model does not treat hardening as a path-dependent
process as the SG model does, but we include it as an indication of the
reported variation of strength with dislocation density or strain state.

\begin{table}
\caption{Plasticity parameters and EOS used for isentrope.}
\label{tab:params}
\begin{center}
\begin{tabular}{|l|l|l|l|l|}\hline
 & $f_v$ & $f_b$ & $M$ & isentrope \\ \hline
Al & fcc & 0.096261 & 3.64 & SESAME 3717 \\
Cu & fcc & 0.123206 & 3.64 & SESAME 3336 \\
Ag & fcc & 0.126552 & 3.64 & LEOS 470 \\
Ir & fcc & 0.111460 & 3.64 & LEOS 770 \\
Pt & fcc & 0.087128 & 3.64 & LEOS 780 \\
Au & fcc & 0.099675 & 3.64 & LEOS 790 \\
Pb & fcc & 0.192771 & 3.64 & LEOS 820 \\
\hline
Fe & bcc & 0.150433 & 3.0 & SNL-SESAME 2150 \\
Ta & bcc & 0.099455 & 3.0 & LEOS 735 \\
W & bcc & 0.125096 & 3.0 & LEOS 740 \\
\hline
Be & hcp & 0.141415 & 2.0 & LEOS 40 \\
Ru & hcp & 0.12 & 2.0 & LEOS 440 \\
\hline\end{tabular}
\end{center}
\end{table}

\subsection{Low pressure}
\begin{figure}
\includegraphics[scale=0.72]{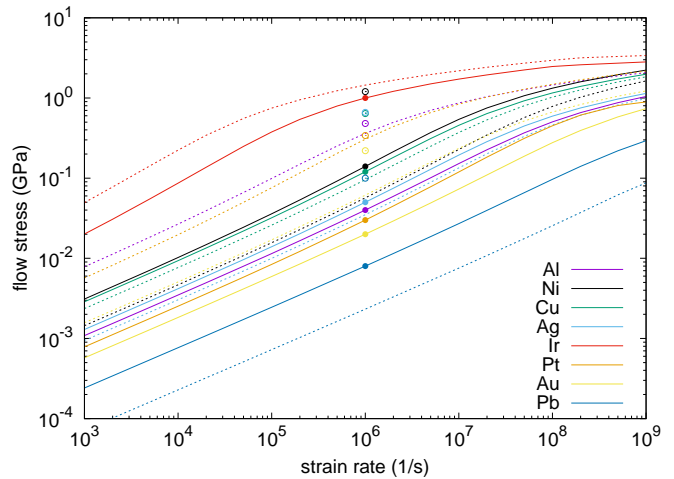}
\caption{Flow stress for face-centered cubic elements at zero pressure,
   predicted with $f_b=0.12$ (dashed) and with $f_b$ fitted to SG $Y_0$ (solid).
   Circles show $Y_0$ (filled) and $Y_{\text{max}}$ (open) where available,
   in the same color as the corresponding line.}
\label{fig:fcclog}
\end{figure}

\begin{figure}
\includegraphics[scale=0.72]{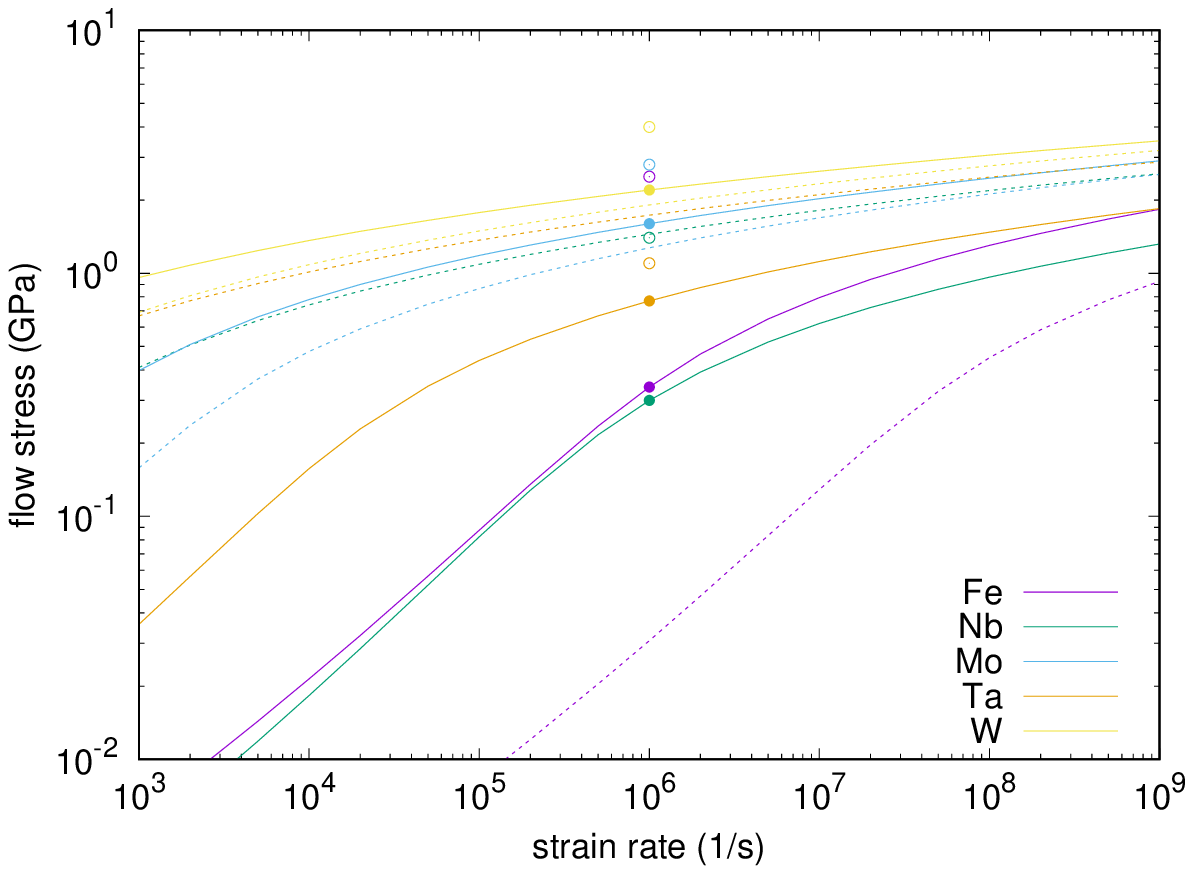}
\caption{Flow stress for body-centered cubic elements at zero pressure,
   predicted with $f_b=0.12$ (dashed) and with $f_b$ fitted to SG $Y_0$ (solid).
   Circles show $Y_0$ (filled) and $Y_{\text{max}}$ (open) where available,
   in the same color as the corresponding line.}
\label{fig:bcclog}
\end{figure}

\begin{figure}
\includegraphics[scale=0.72]{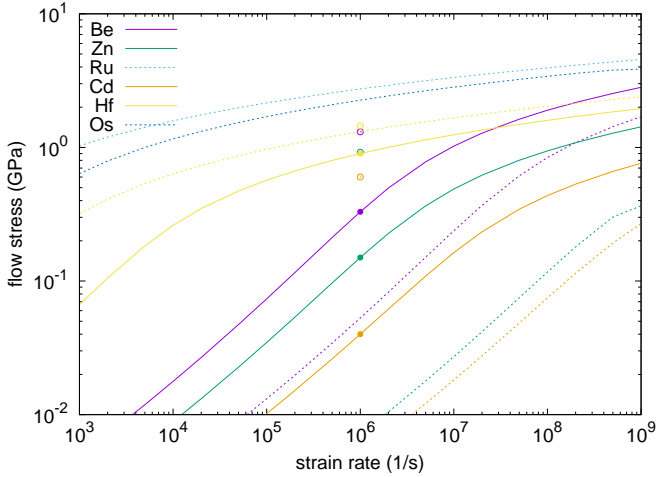}
\caption{Flow stress for hexagonal close-packed elements at zero pressure,
   predicted with $f_b=0.12$ (dashed) and with $f_b$ fitted to SG $Y_0$ (solid).
   Circles show $Y_0$ (filled) and $Y_{\text{max}}$ (open) where available,
   in the same color as the corresponding line.}
\label{fig:hcplog}
\end{figure}

Considering the sensitivity to strain rate at zero pressure
(Figs~\ref{fig:fcclog} to \ref{fig:hcplog}),
taking $f_b=0.12$ gives a prediction that the fcc metals in order of
increasing flow stress are Pb, Ag, Au, Cu, Pt, Al, Ir
whereas the SG $Y_0$ order is Pb, Au, Pt, Al, Ag, Cu (no model for Ir).
It is possible that the SG $Y_0$ may be higher than that of the pure element
where impurities are common, such as Al, Cu, and Pb.
The most striking outlier is Al, which is notable because $f_b=0.12$ appears
valid for melting \cite{Swift_ajmelt_2020}.
For the bcc metals, the predicted order is Fe, Ta, W, in accordance
with the SG $Y_0$. No SG parameter set has been published for pure Fe,
and the fit was performed for stainless steel.
For the hcp metals considered,
Be is predicted to have much lower flow stress than Ru.

For several elements, the $f_b=0.12$ calculation deviated from $Y_0$
by less than the difference between $Y_0$ and $Y_{\text{max}}$,
which we consider encouraging performance for a parameter-free model. 

For most materials starting at the lowest strain rates considered here,
the flow stress is predicted to increase
by roughly a factor of three per order of magnitude increase in strain rate.
At sufficiently high strain rates, the sensitivity is predicted to decrease
as the applied stress reduces the effective height of the Peierls barrier.
This regime is reached at a lower strain rate in stronger elements,
and occurs well below the theoretical maximum strength of the crystal is reached, which would be when $Y=G/2\pi$.
The net effect accounts roughly for the observed difference in elastic wave
amplitudes between microsecond and nanosecond scale experiments.

\begin{figure}
\includegraphics[scale=0.72]{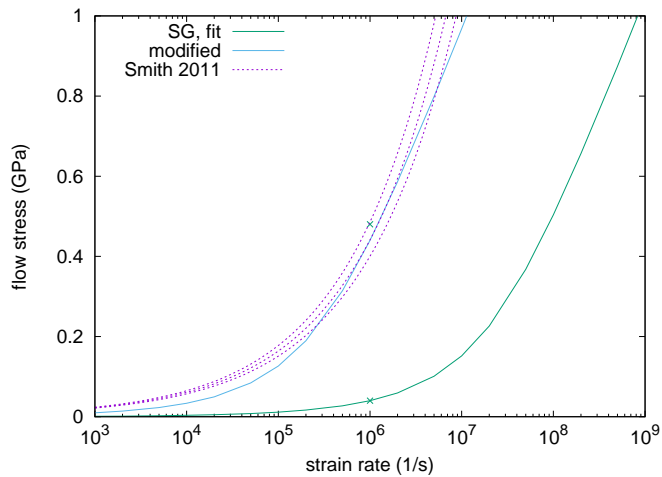}
\caption{Predicted variation of flow stress with pressure and strain rate
   along the principal isentrope of Al and as deduced from rear surface
   velocity measurements \cite{Smith2011}.}
\label{fig:Als}
\end{figure}

The sensitivity of flow stress to relevant strain rates
has previously been investigated in Al and Fe \cite{Smith2011}.
The strain rate was estimated from the rate of acceleration of the rear surface
of the sample, and it was concluded that Al exhibited effects of phonon drag
for strain rates above $10^3$/s, whereas Fe
exhibited a transition from thermal activation to phonon drag above
$\sim 5\times 10^6$/s.
The reduced data exhibited significant scatter in log space, but
there are some inconsistencies in the deduced flow stress, for instance 
the value for Al lay well above the SG $Y_0$ value.
Space- and time-resolved simulations of these experiments capturing the
propagation and evolution of the compression wave may shed more light on
the discrepancy, for example by a modified metric for strain rate
and also by accounting for the evolution from initial dislocation density.
For a simple comparison, we adjusted $f_b$ to reproduce the magnitude
of the fit to the reduced data at $10^6$/s in Al and $5\times 10^6$/s in Fe:
$f_b=$0.122261 and 0.177440 respectively.
Plotting the flow stress in linear space to highlight the stiffness of the 
sensitivity to strain rate, the model prediction for Al reproduces 
the rapid increase in flow stress with strain rate without requiring an
explicit phonon drag term (Fig.~\ref{fig:Als}).
The prediction for Fe did not exhibit the same degree of stiffening.
As well as the same caveats about resolving the propagating waves
and accounting for the evolving dislocation density,
additional complexities for Fe include the impedance mismatch with the
transparent window and phase transformations which occurred during the
laser-driven experiments and could affect the driving stress.

\subsection{High pressure}
We consider the variation of flow stress with pressure along the principal isentrope,
for simplicity in the graphs below neglecting the temperature increase from plastic work,
which is a small correction on the scale of these graphs.
We consider only the fitted value of $f_b$ 
as for most of these metals $f_b=0.12$ did not reproduce the SG $Y_0$
adequately.
Atom-in-jellium EOS are often inaccurate at pressures below a few tenths
of a terapascal, so we took the principal isentropes from
semi-empirical wide-range EOS libraries \cite{sesame,leos},
as listed in Table~\ref{tab:params}.
To emphasize, the plasticity model above is defined in terms of
$\rho$, $T$, $\epsilon_e$, and $\phi_d$: it does not depend explicitly on
pressure, and so the choice of EOS is not an inherent part of the model.
(The shear modulus and Einstein temperature should in principle be consistent
with the EOS, and the Peierls barrier is also related to the EOS though
less directly.)

As well as the predictions of the present model, we also show
the yield stress calculated using the SG model along the same isentrope.
To span all possible loading paths, we plot curves for both $Y_0$ and $Y_{\text{max}}$, i.e. the least and greatest contribution possible from
work-hardening according to that model.

\begin{figure}
\includegraphics[scale=0.72]{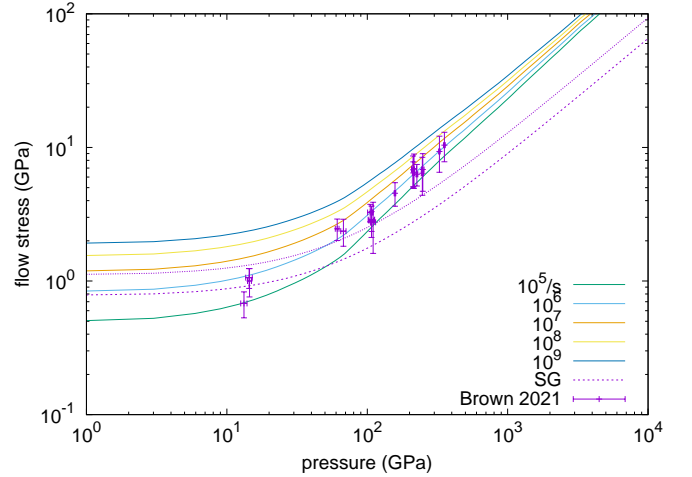}
\caption{Predicted variation of flow stress with pressure and strain rate
   along the principal isentrope of Ta, $f_b$ fitted to SG $Y_0$.
   Magenta lines: SG $Y_0$ (lower) and $Y_{\text{max}}$ (upper), with
   predicted dependence on pressure and temperature along the isentrope.
   Brown data were nominally at $5\times 10^5$/s.}
\label{fig:Ta_isentf}
\end{figure}

\begin{figure}
\includegraphics[scale=0.72]{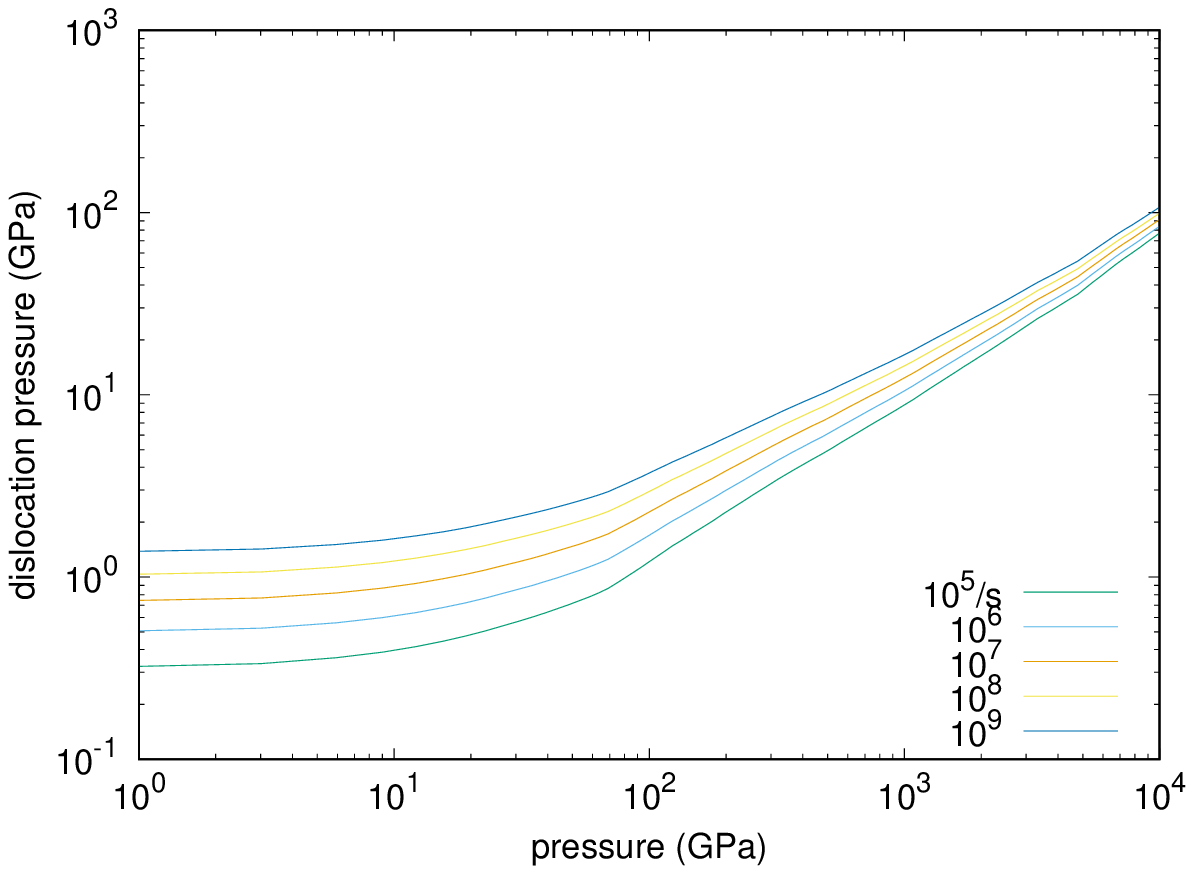}
\caption{Predicted variation of dislocation pressure with pressure and strain rate
   along the principal isentrope of Ta, $f_b$ fitted to SG $Y_0$.}
\label{fig:Ta_isentf_p}
\end{figure}

\begin{figure}
\includegraphics[scale=0.72]{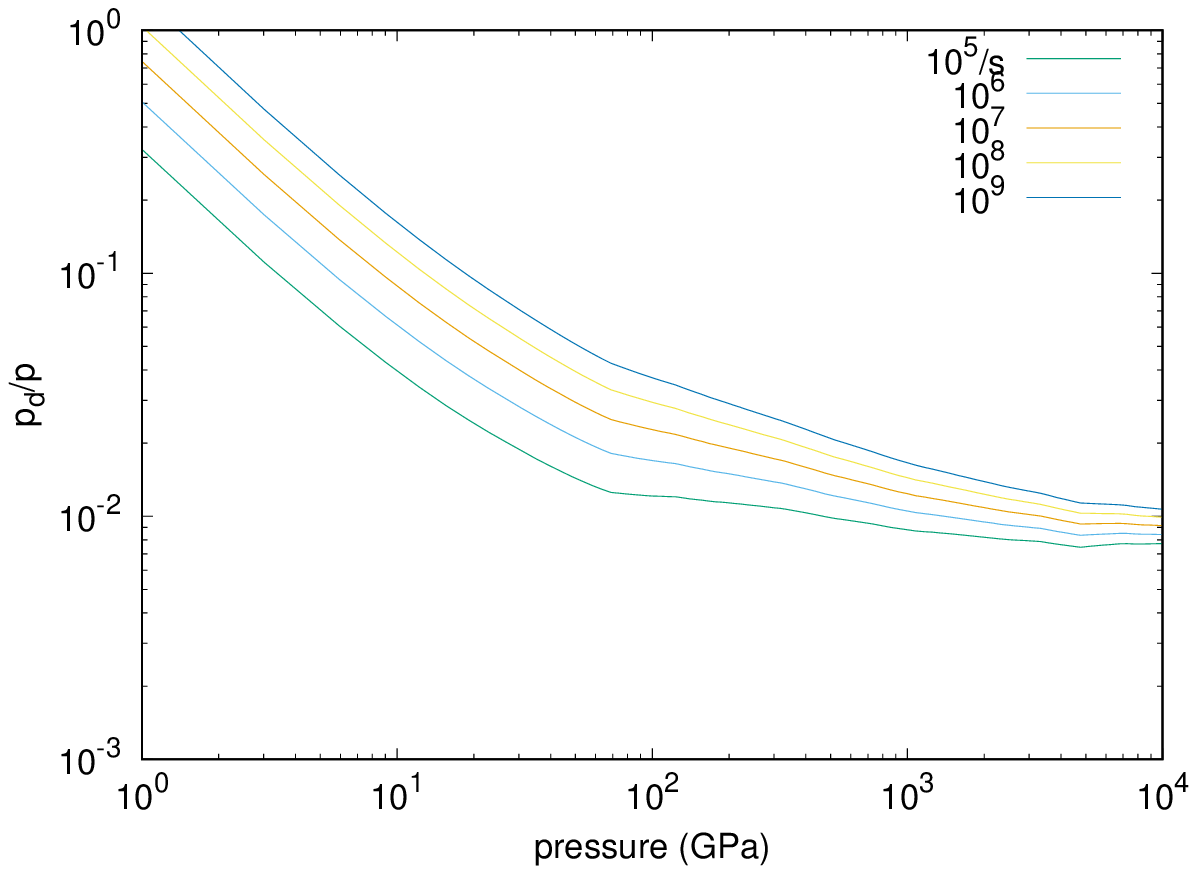}
\caption{Predicted ratio of dislocation pressure $p_d$ to pressure $p$
   along the principal isentrope of Ta, $f_b$ fitted to SG $Y_0$.}
\label{fig:Ta_isenf_fp}
\end{figure}

\begin{figure}
\includegraphics[scale=0.72]{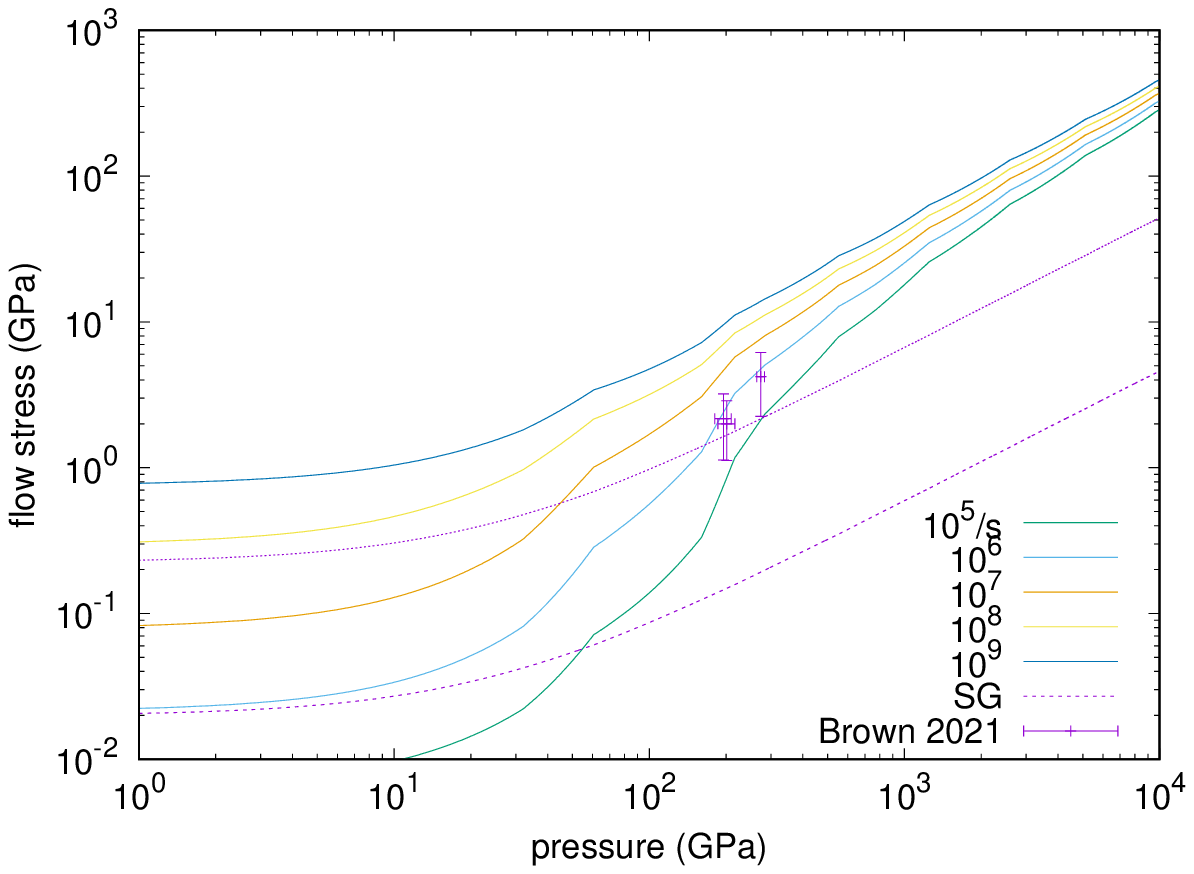}
\caption{Predicted variation of flow stress with pressure and strain rate
   along the principal isentrope of Au, $f_b$ fitted to SG $Y_0$.
   Magenta lines: SG $Y_0$ (lower) and $Y_{\text{max}}$ (upper), with
   predicted dependence on pressure and temperature along the isentrope.
   Brown data were nominally at $5\times 10^5$/s.}
\label{fig:Au_isenf}
\end{figure}

\begin{figure}
\includegraphics[scale=0.72]{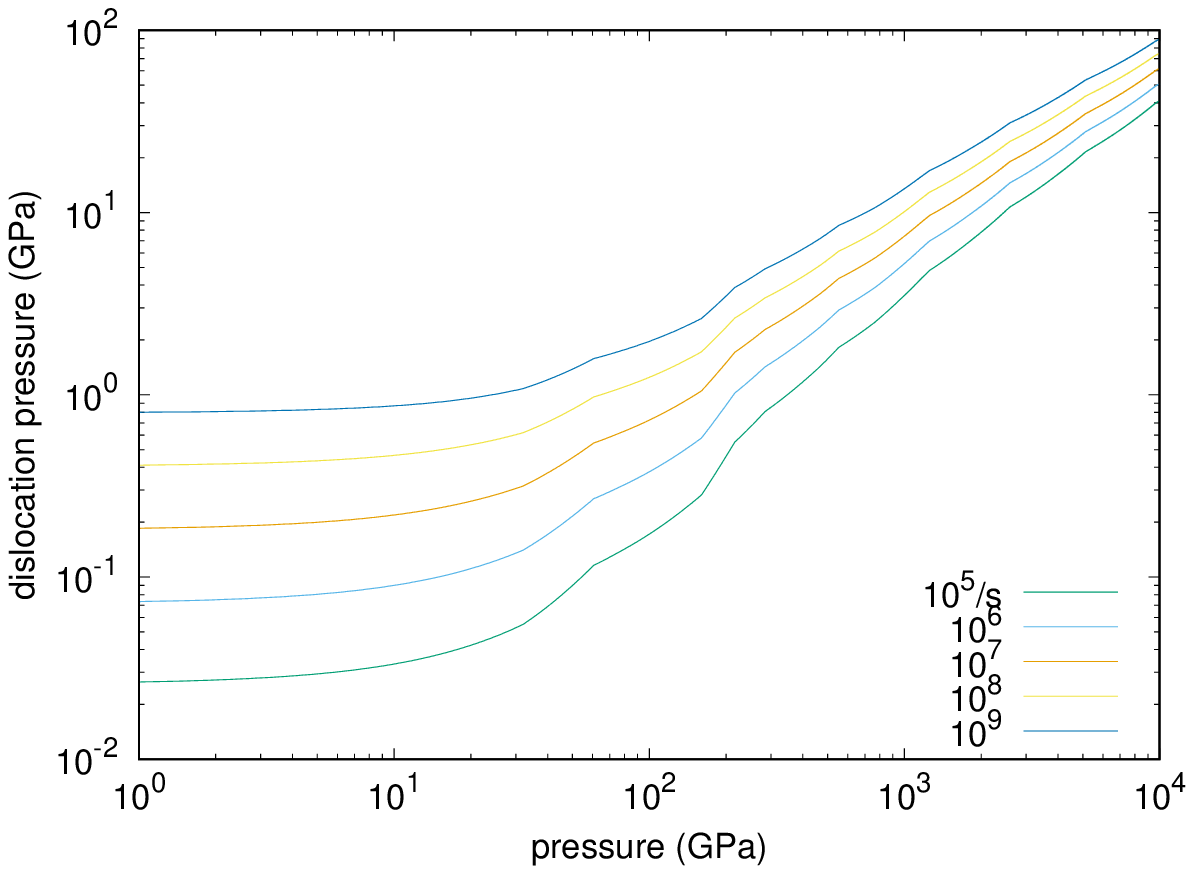}
\caption{Predicted variation of dislocation pressure with pressure and strain rate
   along the principal isentrope of Au, $f_b$ fitted to SG $Y_0$.}
\label{fig:Au_isenf_p}
\end{figure}

\begin{figure}
\includegraphics[scale=0.72]{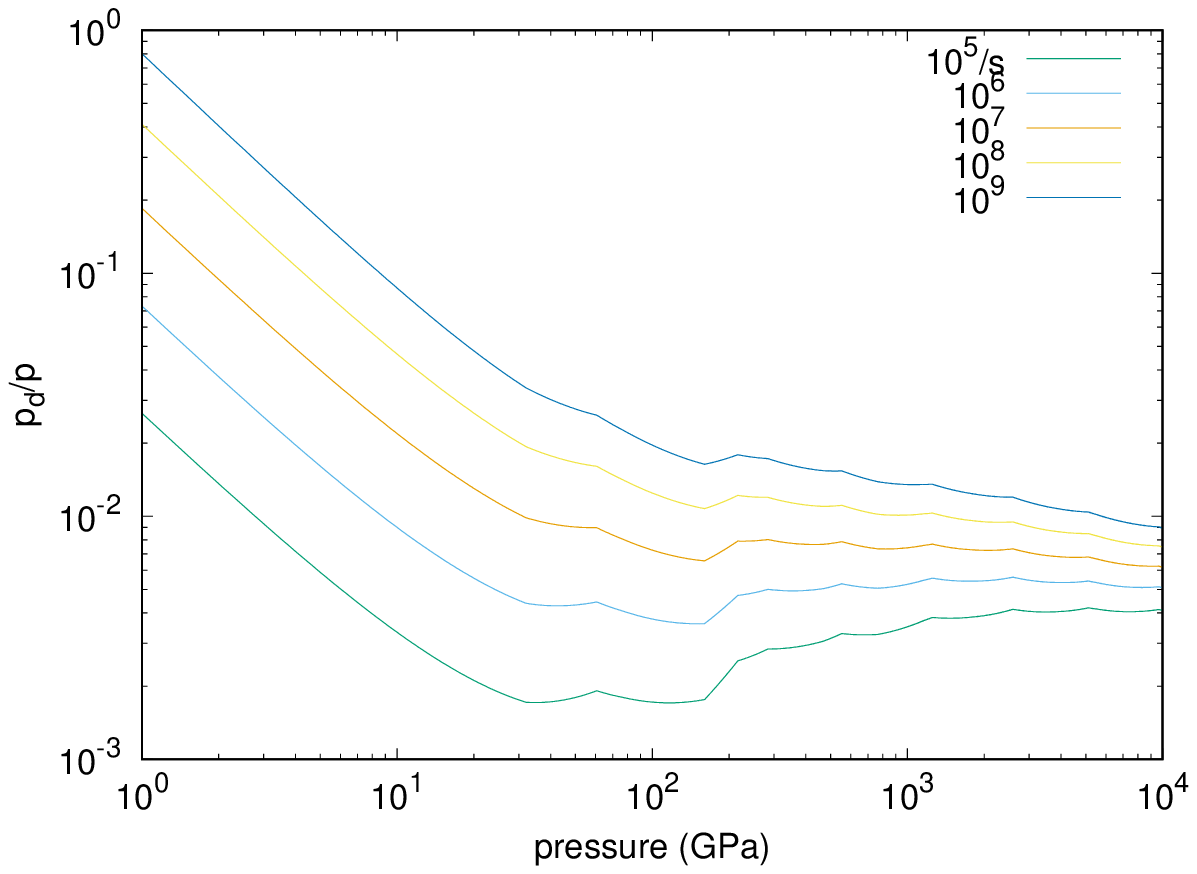}
\caption{Predicted ratio of dislocation pressure $p_d$ to pressure $p$
   along the principal isentrope of Au, $f_b$ fitted to SG $Y_0$.}
\label{fig:Au_isenf_fp}
\end{figure}

\begin{figure}
\includegraphics[scale=0.72]{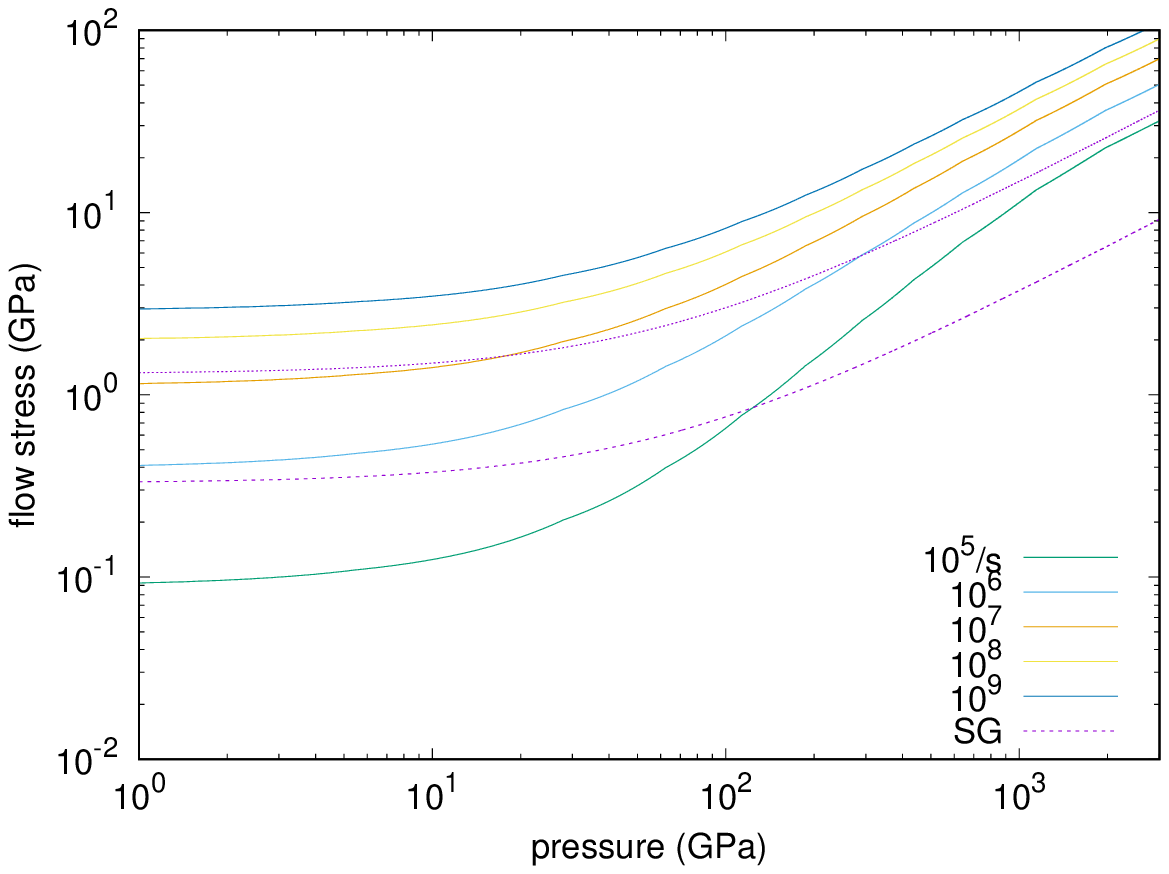}
\caption{Predicted variation of flow stress with pressure and strain rate
   along the principal isentrope of Be, $f_b$ fitted to SG $Y_0$.
   Magenta lines: SG $Y_0$ (lower) and $Y_{\text{max}}$ (upper), with
   predicted dependence on pressure and temperature along the isentrope.}
\label{fig:Be_isenf}
\end{figure}

Considering the behavior in more detail for representative elements
(Figs~\ref{fig:Ta_isentf} to \ref{fig:Be_isenf}), those of lower $Z$
such as Be are predicted to retain a large dependence of flow stress on
strain rate to high pressures.
For fcc elements of moderate to high $Z$,
the sensitivity of flow stress to strain rate is large
at low pressures but decreases at pressures above $\sim 100$\,GPa.
For bcc elements of moderate to high $Z$,
the sensitivity of flow stress to strain rate is less than that of fcc 
elements at low pressure, but decreases more slowly with pressure.
For Ta in particular, the flow stress at pressures of a few hundred gigapascals
and a strain rate $10^7$/s
representative of Rayleigh-Taylor ripple growth experiments is 
close to twice the SG $Y_{\text{max}}$ value\footnote{%
Rayleigh-Taylor ripple growth experiments are typically designed to give
a plastic strain of a few tens of percent.
At these strains, the work-hardening term in the SG strength model for Ta
has reached $Y_{\text{max}}$.
}.
The flow stress of metals including Ta and Au has been estimated
at pressures $\sim 0.1$\,TPa
from release features in the velocity history transmitted through ramp-loaded samples
\cite{Brown2021}.
The strain rate in these experiments was nominally $5\times 10^5$/s,
but it was likely somewhat lower at low pressures and higher at high pressures,
and we suggest that the measurements likely sampled a range $10^5-10^6$/s.
This is a recent experimental development and the uncertainties are relatively large,
but the trend and magnitude appear consistent with our predictions.
\condbf{%
A recent study \cite{Prime2022} found that previous multi-scale plasticity models
and models constructed using more detailed dislocation processes do not
accurately predict the flow stress at pressures above $\sim$0.1\,TPa,
and require further study.
Specifically, the models evaluated for Ta were not consistent with
experimental measurements at higher pressure unless recalibrated to match.
It is striking that the present theory predicts flow stresses consistent
with the measurements when calibrated to a single, low-pressure datum.
Either the dominant mechanisms of plastic flow do not change between
the fitting and test states,
or any competing mechanism fortuitously exhibits the same dependencies.
}

The contribution of the core and strain energy of the dislocations
is predicted to be significant.
At low pressures, calculated as above, it can easily exceed the pressure from
the EOS.
This is misleading, because we are concerned with near-uniaxial loading,
and it is not possible to sustain the strain rate for long enough to build up
an equilibrium population of dislocations before the pressure rises
substantially.
However, for pressures into the terapascal range, the dislocation pressure
tends toward a few percent of the mean pressure, which may be a significant
correction when interpreting dynamic loading behaviour in terms of
the scalar EOS.

\section{Conclusions}
We describe a continuum-level plasticity model for polycrystalline materials
based on a single dislocation density and single mobility mechanism,
with an evolution model for the dislocation density.
Using the approximate but wide-ranging and computationally efficient
atom-in-jellium model for the electronic structure of matter to
predict the shear modulus, Einstein frequency, and systematic variation of the
Peierls barrier, we deduce a credible variation of flow stress
with pressure and strain rate.
In this formulation,
the equilibrium dislocation density is an emergent property of the model.
With the dislocation density expressed per-atom rather than as a dislocation
length per volume, to simplify the plasticity relations at finite
compressions, its value was predicted to vary less with pressure.
Interestingly, stiffening at high strain rates that is usually attributed
to phonon drag may be accounted for by the form of the hopping rate.

The model is constructed to make it straightforward to substitute alternative
forms for key components, such as the shear modulus, hopping attempt rate,
Peierls barrier, and terms in the dislocation evolution equation,
where more accurate relations are available.
It may also be possible to generalize the dislocation mobility to account
for several types of barrier, represented with a rheonet.

In estimating the Peierls barrier from the Einstein frequency,
we find a potential link to the Lindemann melting law,
which leads to a parameter-free approximate plasticity model.
This approach otherwise requires a single free
parameter to be determined from experiment or more detailed theory.

Accounting for the elastic energy of the dislocations gives a prediction of
the plastic work appearing as heat, equivalent to a dynamic, self-consistent
estimate of the Taylor-Quinney factor,
which has long been a poorly-constrained parameter or assumption in 
simulations of plastic flow.
The elastic energy of the dislocations is predicted to be manifested also
as a contribution to the pressure, 
which may reach several percent of the total pressure
in ramp compression to the terapascal range, and should therefore be
considered when interpreting ramp data as an isentrope or isotherm.
The bulk elastic shear energy also contributes to the pressure,
but is much smaller.
The mean pressure is usually assumed to depend only on
the scalar equation of state and not in general on the elasticity or strength: 
both of these contributions represent new physics
at high dynamic pressures.
We also find that a constant shear modulus would lead to a {\it negative}
contribution to pressure from both elastic shear and dislocation energies,
which may not have been previously appreciated.

The model reproduces the observation that the elastic wave amplitude
for loading of samples $o(10)$\,$\mu$m thick, on nanosecond time scales,
corresponds to a flow stress a few times higher than for samples $o(10)$\,mm
thick, on microsecond time scales.
It predicts a reduced dependence of flow stress on strain rate
in higher-$Z$ materials, particularly in the face-centered cubic structure,
at pressures over $\sim 100$\,GPa.
The predictions are reasonably consistent with recent assessments of
strength at high pressure from surface velocimetry data,
\condbf{%
whereas other known plasticity models were not consistent without
calibration to the data.
}

\section*{Acknowledgments}
The authors would like to acknowledge informative conversations with
Dean Preston, Vasily Bulatov, Nicolas Bertin, Jared Stimac, Sylvie Aubry,
Tom Arsenlis, Nathan Barton, Robert Rudd, and Philip Sterne.
The anonymous journal referees also made valuable suggestions.
This work was performed under the auspices of
the U.S. Department of Energy under contract DE-AC52-07NA27344.

\appendix
\section*{Appendix A: Pressure contributions}
The contributions to specific energy of the material 
from the dislocation population and from the bulk elastic 
shear strain take the form
\begin{equation}
e\propto G/\rho = \alpha G/\rho \text{, say}.
\label{appeq:en}
\end{equation}
Pressure depends on the Helmholtz specific free energy $f$ as
$p=-\partial f/\partial v|_T$
where $v$ is the specific volume, $1/\rho$,
so
\begin{equation}
p=\rho^2\left.\diffl f\rho\right|_T.
\end{equation}
As usual for EOS construction, we assume that
contributions to free energy are additive,
which means that the corresponding contributions to pressure
can be also considered separately and added.
The shear modulus $G$ varies in principle with $\rho$ and $T$.
Its temperature variation subsumes the implicit entropy dependence of $G$.
If we neglect any explicit entropy dependence of these contributions
to the specific energy then $f=e$,
which is equivalent to assessing the pressure contributions along the cold
compression curve.
Differentiating,
\begin{equation}
p=\alpha\left(\rho\left.\diffl G\rho\right|_T-G\right).
\end{equation}
Factorizing out $G$ and noting that $\alpha G=\rho e$ by Eq.~\ref{appeq:en},
we find that the pressure contribution can be expressed with respect to
the dimensionless logarithmic derivative of $G$ with respect to $\rho$:
\begin{equation}
p=\rho e\left(\frac \rho G\left.\pdiffl G\rho\right|_T-1\right).
\end{equation}

This derivation implies that a constant shear modulus leads to negative
pressure contributions from the dislocations and shear strain, 
which may seem surprising.
However, it follows by inspection from Eq.~\ref{appeq:en}:
increasing $\rho$ at constant $G$ leads to a decrease in specific energy,
and thus leads to a tensile contribution to pressure.
If we consider a power-law dependence $G\propto\rho^\beta$ then
\begin{equation}
p=\rho e(\beta-1).
\end{equation}
Thus, in order for these pressure contributions to be positive,
the effective exponent for the dependence of $G$ on $\rho$ must exceed unity.

\section*{Appendix B: Deviatoric elastic strain}
The contribution of deviatoric elastic strain to pressure is a minor part of
the results reported, but caused some discussion during manuscript review.

Studies of quasistatic loading often employ high-order elasticity,
which expresses the components of the stress tensor as a Taylor expansion
about the initial state.
Third-order components can capture a connection between shear and pressure,
and this has been recognized in low-symmetry crystal structures at low
pressure and also in soft materials exhibiting large elastic strains,
such as rubbers.

The analysis presented here implies that elastic shear in a material
described even by only second-order deviatoric elastic moduli results in a
contribution to pressure, depending on the variation of the moduli
with compression.
This result has been deduced previously in a similar though not identical form, 
following a longer derivation \cite{Scheidler1996} which also considers
the effect of third-order contributions.
The previous analysis is equivalent to ours, but there is an error
in the associated discussion that may have caused readers some confusion:
Eq.~4.7 in the report reads
$$
p_2>0\iff
\tilde\rho \diffl\mu{\tilde\rho}>\mu\iff
\diffl\mu{\tilde p}>\frac\mu\kappa\iff
\diffl{U_s}{\tilde p}>0
$$
where
$p_2$ is the contribution to isotropic pressure $\tilde p$,
$\tilde\rho$ is the compression ratio with respect to the initial state $=\rho/\rho_0$,
$\mu$ is the shear modulus, $\kappa$ the bulk modulus, and
$U_s$ the shear wave speed.
The inequalities are equivalent to testing $(\rho/G) \partial G/\partial\rho>1$
in our Eq.~\ref{eq:pe}.
In the discussion immediately following Eq.~4.7, the report states
``$dU_s/d\tilde p$ implies $d\mu/d\tilde p>0$'' which is inconsistent with
Eq.~4.7 and with our derivation, though the report goes on to use Eq.~4.7
consistently to evaluate pressure-shear coupling at ambient conditions in
seven materials including four elemental metals.
It seems possible that an error such as this may have led to the misconception
that a constant shear modulus leads to no pressure-shear coupling,
as this report has been cited in publications in this area.

Pressure-shear coupling has been linked with anisotropy
\cite{Fuller2011} in a literature that indicates that such phenomena are
recognized for geologic materials and ceramics,
but not for metals because elastic strains tend to be too small for anisotropic
effects to become significant \condbf{or do not occur in cubic crystals}.
This narrative is consistent with our work, because dynamic loading under
HED conditions can induce high compressions at relatively low temperatures,
on time scales where the elastic strain may be much larger than customary
and thus potentially important to account for.
\condbf{It has also been suggested that pressure-shear coupling can only
occur in non-cubic crystals \cite{Luscher2017}.}
However, pressure-shear coupling in our analysis and that of \cite{Scheidler1996}
may occur from purely isotropic deviatoric elasticity,
\condbf{and may occur in cubic crystals, although of course they cease to be
cubic when deformed}.

Pressure-shear coupling can also be captured in high-order elasticity theory,
in which the stress tensor is expressed as a Taylor series with respect to
elastic strains from a reference state, usually the initial, undeformed state.
Pressure-shear contributions can occur from third-order and higher terms,
in contrast to the contribution from second-order deviatoric elasticity
that occurs here via the compression dependence of the shear modulus.
\condbf{%
Although dislocation plasticity models have been constructed
on higher-order elasticity theory
\cite{Nguyen2021},}
this formalism is not an appropriate over the range of pressures of interest to us.
Matter may exhibit orders of magnitude of isotropic compression,
whereas the maximum elastic shear strain possible is a few tens of percent
(depending on on crystal structure and shear orientation) before the
shear stress decreases by symmetry of the lattice.
Therefore, a Taylor expansion in strain from the initial state 
is less appropriate than a strain-invariant isotropic model such as a
scalar EOS, supplemented by a Taylor expansion about the state of isotropic
stress.

\vfill

\end{document}